\documentclass[twocolumn]{aastex63}
\usepackage{amsmath}
\usepackage{mathtools}
\usepackage{amssymb}
\usepackage{lineno}

\shorttitle{GRB 211211A from NSWD Merger}
\shortauthors{Zhong et al.}

\begin{document}
\title{GRB 211211A: a Neutron Star$-$White Dwarf Merger?}
\author[0000-0002-1766-6947]{Shu-Qing Zhong}
\affil{Deep Space Exploration Laboratory / Department of Astronomy, University of Science and Technology of China, Hefei 230026, People's Republic of China; sqzhong@ustc.edu.cn, daizg@ustc.edu.cn}
\affil{School of Astronomy and Space Science, University of Science and Technology of China, Hefei 230026, People's Republic of China}
\author[0000-0002-8391-5980]{Long Li}
\affil{Deep Space Exploration Laboratory / Department of Astronomy, University of Science and Technology of China, Hefei 230026, People's Republic of China; sqzhong@ustc.edu.cn, daizg@ustc.edu.cn}
\affil{School of Astronomy and Space Science, University of Science and Technology of China, Hefei 230026, People's Republic of China}
\author[0000-0002-7835-8585]{Zi-Gao Dai}
\affil{Deep Space Exploration Laboratory / Department of Astronomy, University of Science and Technology of China, Hefei 230026, People's Republic of China; sqzhong@ustc.edu.cn, daizg@ustc.edu.cn}
\affil{School of Astronomy and Space Science, University of Science and Technology of China, Hefei 230026, People's Republic of China}
\affil{School of Astronomy and Space Science, Nanjing University, Nanjing 210023, People's Republic of China}

\begin{abstract}
The gamma-ray burst GRB 211211A and its associated kilonova-like emission were reported recently.
A significant difference between this association event and GRB 170817A/AT 2017gfo is that GRB 211211A has a very long duration.
In this paper, we show that this association event may arise from a neutron star$-$white dwarf (NS$-$WD) merger if a magnetar leaves finally in the central engine.
Within the NS$-$WD merger, the main burst of GRB 211211A could be produced by magnetic bubble eruptions from toroidal magnetic field amplification of the pre-merger NS.
This toroidal field amplification can be induced by the runaway accretion from the WD debris disc if the disc is in low initial entropy and efficient wind. 
While the extended emission of GRB 211211A is likely involved with magnetic propelling. The observed energetics and duration of the prompt emission of GRB 211211A can be fulfilled in comparison with those of accretion in hydrodynamical thermonuclear simulation, as long as the WD has a mass $\gtrsim1M_{\odot}$.
Moreover, if the X-ray plateau in GRB afterglows is due to the magnetar spin-down radiation, GRB optical afterglows and kilonova-like emission can be well jointly modeled combining the standard forward shock with the radioactive decay power of $^{56}{\rm Ni}$ adding a rotational power input from the post-merger magnetar.
\end{abstract}

\keywords{Compact binary stars (283); Gamma-ray bursts (629); Magnetars (992); White dwarf stars (1799)}

\section{Introduction}
\label{sec:introduction}
The milestone event of gravitational wave (GW)/short gamma-ray burst (SGRB)/kilonova association
\citep[GW170817/SGRB 170817A/AT 2017gfo;][]{abbott17c,abbott17a,abbott17b,gold17,sav17,cou17,arc17,lip17,kas17,soa17,tan17,val17,ale17,evans17,hall17,marg17,tro17,drout17,cho17,nicholl17a,pian17,sha17,sma17}
firstly confirmed that neutron star$-$neutron star (NS$-$NS) merger at least is one source of SGRBs
and kilonovae, the detail can be referred to the review \citep{mar21}.
Before and after this event, there was no another confirmed kilonova-associated SGRB,
even a few kilonova candidate bursts such as SGRBs 050709 \citep{jin16},
050724A \citep{gao17}, 060614 \citep{yang15}, 070714B \citep{gao17}, 070809 \citep{jin20},
130603B \citep{berger13,tanvir13}, 150101B \citep{tro18,gom18}, 160821B
\citep{kasli17b,lamb19,tro19}, and
200522A \citep{fong21,ocon21}.

Until recently, a kilonova-like emission with similar luminosity, duration and color to AT 2017gfo,
being associated with GRB 211211A and locating in a nearby host galaxy SDSS
J140910.47+275320.8 with a distance $\sim350$ Mpc was reported and analyzed by \cite{ras22}, \cite{yang22}, \cite{mei22}, \cite{tro22}, \cite{gom23}, and \cite{xiao22}.
From the evidence of the host galaxy properties including the offset, the features of the kilonova-like emission, 
the modeling for the light curves of afterglows and kilonova-like emission,
and the exponential decline phase and spectral features in GRB 211211A,
most of authors suggested or directly considered this association to arise from a compact star merger 
\citep{ras22,yang22,tro22,gom23,mei22,zhang22,xiao22,gao22,zhu22,chang23,kun23},
while a few authors mainly discussed the explanation or possibility in a collapsar origin \citep{wax22,barn23}.
The final central engine either a BH or magnetar is also under debate.
More notably, the accompanying GRB 211211A with an extended emission (EE-SGRB) lasts a minute long duration,
resembling those EE-SGRBs originating from a compact star merger rather than a massive star core-collapse
such as GRBs 050724A \citep{norris06}, 060505 \citep{ofek07}, 060614 \citep{gal06,della06,fynbo06,geh06,met08,buc12},
GRB 150424A \citep{knu17}
GRB 211227A \citep{lv22}, and etc \citep{norris10,gom13,gom20,kis17,lan20}.
Nonetheless, it should be noted that GRB 211211A has a long main burst (MB) lasting $\gg2$ s,
significantly longer than those EE-SGRBs with an initial short spike mentioned above.
This motivates us to think that the EE-SGRBs with a long-duration MB, for example, GRBs 211211A and 060614, are likely from an NS$-$WD merger,
differing from an NS$-$NS or NS$-$BH merger usually applied to account for those EE-SGRBs with a short MB $\lesssim2$ s.

An NS$-$WD merger was used to explain GRB 211211A by \cite{yang22}, who proposed that
a newly-born magnetar would form through the neutronization of the WD materials and its mixture with the materials from the pre-merger NS, 
during the collapsing of the WD induced by the NS merging into the center of the WD.
In their model, both the MB and EE of GRB 211211A are considered to be likely from differential-rotation-induced magnetic bubbles of the newly-born magnetar, the X-ray plateau in afterglows is attributed to the magnetar wind due to the spin-down, and the kilonova-like emission is a kilonova powered by neutron-rich ejecta and magnetar rotational energy injection.
In this paper, nevertheless, we propose another interpretation for
the observations of GRB 211211A in the framework of an NS$-$WD merger.
In our model, there are at least two aspects distinguishing from the model of \cite{yang22}: (1) the MB of GRB 211211A is generated through the magnetic field amplification of the pre-merger NS induced by the material accretion from the tidally disrupted WD, while its EE is yielded by the magnetic propelling due to the late accretion; (2) the kilonova-like emission is regarded as a rapidly evolving transient (RET) powered by $^{56} \rm{Ni}$ decay adding an energy injection from the spin-down of the post-merger magnetar.

Hereafter, we show that the prompt emission (MB and EE) of GRB 211211A (\S \ref{sec:prompt}), its afterglows and kilonova-like emission (\S \ref{sec:kilo-after})
all can be interpreted within the framework of an NS$-$WD merger
if a magnetar leaves in the central engine postmerger.
Throughout the paper, the notation $Q_n=Q/10^n$ in cgs is adopted.

\section{Prompt Emission of GRB 211211A}
\label{sec:prompt}
In general, there are two evolutionary pathways for an NS$-$WD binary,
depending on its WD mass $M_{\rm WD}$ and the critical WD mass $M_{\rm WD,crit}$.
The first pathway is that the NS$-$WD binary undergoes
a stable mass transfer then evolves into
an ultra-compact X-ray binary if $M_{\rm WD}<M_{\rm WD,crit}$.
The second pathway is that the NS$-$WD binary experiences an unstable mass transfer then enters a merger phase
in a rather short dynamical timescale for $M_{\rm WD}>M_{\rm WD,crit}$ \citep{hjel87,hurley02}.
The critical WD mass is found
to be $M_{\rm WD,crit}=0.37~M_{\odot}$ \citep{van12}
or a lower one $0.2~M_{\odot}$ \citep{bob17}.
\cite{too18} suggested that over 99.9\% of semi-detached NS$-$WD binaries
would merge when $M_{\rm WD,crit}=0.2~M_{\odot}$,
which indicates that the merger is a prevalent fate for semi-detached NS$-$WD binaries.

The unique type of long GRBs without an associated supernova (SN) like the remarkable case GRB 060614,
was firstly suggested to be produced by the accretion from the disrupted WD debris in an NS$-$WD merger by \cite{king07}.
From the statistical evidence in the fluence, duration, peak energy of $\nu f_{\nu}$ spectrum, and isotropic energy distributions for the BATSE GRB catalog,
\cite{chatt07} came to a similar conclusion that one type of GRBs could invoke an NS$-$WD merger.
The questions are what possible mechanism can generate a GRB during the accretion in an NS$-$WD merger
and how energetic the GRB can reach.

{\em What mechanism can generate a GRB?} One possible mechanism is differential-rotation-induced magnetic bubbles \citep{kluz98,ruder00,dai06}
if the toroidal magnetic field of the pre-merger NS can be amplified due to a runway accretion \citep{zhong20}.
The basic picture is somewhat different from that illuminated in \cite{yang22}:
due to the differential rotation between the inner core and the outer shell of the pre-merger NS induced by the accretion from the WD debris to the surface of the NS,
the toroidal magnetic field $B_{\phi}$ of the NS can be gradually amplified via the windup process
and forms a magnetically confined toroid.
Once the strength of $B_{\phi}$ exceeds the buoyancy strength $B_{\rm b}\sim10^{17}$ G \citep{kluz98},
the magnetic toroid penetrates through the surface of the NS and may rapidly reconnect typically within $10^{-4}$ s to yield a magnetic bubble eruption.
Continuous magnetic bubbles would be produced by continuous $B_{\phi}$ amplification because of continuous accretion.
One significant difference between our model and the model in \cite{yang22} is the source of the differential rotation.
For a proto NS born from the collapse of the WD in their model, the differential rotation actually comes from the pre-collapse WD itself and the potential interaction between the WD and the NS.
While for a ``renascent" NS in our model, the differential rotation originates from the angular momentum and energy
transport of the accreting WD debris onto the NS \citep[see Footnote 5 of][]{zhong20}.

{\em How energetic can the GRB reach?} In any case, in our model the $B_{\phi}$ amplification and the subsequent magnetic bubble eruption are like intermediary processes between the runway accretion
and the GRB generation,
so the GRB luminosity $L_{\rm j}$ can be directly related to the accretion power, i.e.,
\begin{equation}
L_{\rm j}= \epsilon_{\rm j} \dot{M} c^{2}\simeq 1.8\times10^{51}\ \epsilon_{\rm j,-1}\dot{M}_{-2}\ {\rm erg~s^{-1}},
\label{eq:L_j}
\end{equation}
where $\epsilon_{\rm j,-1}=\epsilon_{\rm j}/10^{-1}$ is the efficiency of the accretion power
converting to the GRB power and $\dot{M}_{-2}=\dot{M}/10^{-2}M_{\odot}{\rm s}^{-1}$ is the accretion rate.
For GRB 211211A, its peak luminosity of the whole burst is $1.9\times10^{51}~{\rm erg~s^{-1}}$,
which can be reached as long as the peak accretion rate can be $\gtrsim0.01M_{\odot}{\rm s}^{-1}$.
As shown in Figure 3 of \cite{kal22} for an NS$-$WD binary with mass \footnote{Note that the properties of accretion and wind ejection, for instance, the total accretion mass, peak accretion rate, accretion duration, etc., between a $M_{\rm NS}:M_{\rm WD}=1.25:1.00$ binary and a $M_{\rm NS}:M_{\rm WD}=1.40:1.00$ binary are comparable, see Table 4 of \cite{kal22}. Thus we adopt a $M_{\rm NS}:M_{\rm WD}=1.40:1.00$ binary throughout the paper.} $M_{\rm NS}:M_{\rm WD}=1.25:1.00$ in units of $M_{\odot}$, the peak accretion rate can be $\gtrsim0.01M_{\odot}{\rm s}^{-1}$ if the accretion disk is in low initial entropy and efficient wind, which is larger than that in high initial entropy\footnote{The accretion rate in an NS$-$WD merger was studied in high initial entropy in all previous works such as \cite{mar16} and \cite{fern19}, so its peak value never reaches $10^{-3}M_{\odot}{\rm s}^{-1}$.} by about two orders of magnitude.
Moreover, also seen in \cite{kal22}, the accretion rate $\gtrsim10^{-3}M_{\odot}{\rm s}^{-1}$ [the average luminosity of the whole burst of GRB 211211A is $\sim10^{50}~{\rm erg~s^{-1}}$ corresponding to an accretion rate $\sim10^{-3}M_{\odot}{\rm s}^{-1}$ using Equation (\ref{eq:L_j})] lasts a few tens seconds, which is comparable to the duration of the whole burst of GRB 211211A.
Further, the isotropic radiation energy of the whole burst $E_{\gamma,\rm iso}\simeq7.6\times10^{51}~{\rm erg}$ is just about 2\% of that from the total accretion material $E_{\rm acc}=M_{\rm acc}c^2\simeq4.3\times10^{53}M_{\rm acc,0.24}~{\rm erg}$ (here the accretion mass $M_{\rm acc}=0.24M_{\odot}M_{\rm acc,0.24}$ is adopted from Table 4 of \cite{kal22} for a $M_{\rm NS}:M_{\rm WD}=1.40:1.00$ binary).
As a result, the energetics and duration of the whole burst of GRB 211211A can be well satisfied in an NS$-$WD binary with $M_{\rm NS}:M_{\rm WD}=1.40:1.00$.
But it should be noted that one implicit requirement is the toroidal field $B_{\phi}$ of the NS can be quickly amplified to the buoyancy strength $B_{\rm b}\sim10^{17}$ G within a timescale
much less than the accretion timescale.
Next we will address this issue.

{\em Performing Numerical Calculation.} Back to the $B_{\phi}$ amplification
due to the accretion,
we assume that the NS is divided as the core and the shell by a boundary at radius $R_{\rm c}\cong 0.5R$ in which $R$ is the radius of the NS. 
Moreover, the moments of inertia of the core and the shell are adopted as $I_{\rm c}\cong I_{\rm s}=0.5I$ \citep{spr99,dai06}, 
in which $I=\frac{2}{5}M_{\rm NS}R^2$ 
is the total moment of inertia of the star.
Though most of equations in performing are documented in \cite{zhong20},
we briefly rewrite them here for more general situations.

The accretion invokes three critical radii.
The first one is the Alfv\'{e}n radius
\begin{eqnarray}
r_{\rm m}&=&(B_{\rm s} R^3)^{4/7}(GM_{\rm NS})^{-1/7}\dot{M}^{-2/7} \nonumber\\
&\simeq& 28\ B_{\rm s,15}^{4/7}R_{6}^{12/7}M_{\rm NS,1.4}^{-1/7}\dot{M}_{-3}^{-2/7}\ {\rm km},
\label{eq:r_m}
\end{eqnarray}
where $B_{\rm s,15}=B_{\rm s}/10^{15}{\rm G}$, $R_{6}=R/10^6{\rm cm}$, and $M_{\rm NS,1.4}=M_{\rm NS}/1.4M_{\odot}$ are
the surface dipole magnetic field strength, radius, and mass of the NS, respectively.
The second one is the corotation radius
\begin{eqnarray}
r_{\rm c}=\left(\frac{GM_{\rm NS}}{\Omega_{\rm s}^2}\right)^{1/3}
\simeq7.8\times10^3M_{\rm NS,1.4}^{1/3}P_{\rm s,1}^{2/3}\ {\rm km},
\label{eq:r_c}
\end{eqnarray}
where $\Omega_{\rm s}=2\pi/P_{\rm s}$ and $P_{\rm s,1}=P_{\rm s}/10{\rm s}$
are the angular velocity of the shell and its relative spin period. These two radii determine the accretion form: for $r_{\rm m}<r_{\rm c}$, material is funneled by the NS dipole field and  accreted onto the NS surface, the NS spins up; while for $r_{\rm m}<r_{\rm c}$, material must spin at a super-Keplerian rate to come into corotation with the NS and is thus expelled, the NS spins down \citep{ill75}.
The third one is the light cylinder radius,
\begin{eqnarray}
r_{\rm lc}=c/\Omega_{\rm s}\simeq4.8\times10^5P_{\rm s,1}\ {\rm km}.
\label{eq:r_lc}
\end{eqnarray}

During the accretion, the
time-dependent angular velocities for the shell ($\Omega_{\rm s}$) and the core ($\Omega_{\rm c}$) of the NS can be solved by
\begin{eqnarray}
	I_{\rm s}\frac{d\Omega_{\rm s}}{dt} = N_{\rm acc}-N_{\rm dip}-N_{\rm mag},
	\label{eq:I_s}
\end{eqnarray}
and \citep{spr99}
\begin{eqnarray}
	I_{\rm c}\frac{d\Omega_{\rm c}}{dt} = N_{\rm mag},
	\label{eq:I_c}
\end{eqnarray}
where $N_{\rm acc}$, $N_{\rm dip}$, and $N_{\rm mag}$ are the accretion, magnetic dipole, and magnetic torques, respectively.
The first torque is written by \citep[e.g.,][]{piro11}
\begin{equation}
N_{\rm acc}=\begin{cases}\left(1-\frac{\Omega_{\rm s}}{\Omega_{\rm K}}\right)(GM_{\rm NS}R)^{1/2}\dot{M},
&r_{\rm m}\leqslant R, \\
n(\omega)\left(GM_{\rm NS}r_{\rm m}\right)^{1/2}\dot{M}, &r_{\rm m}>R,
\end{cases}
\label{eq:N_acc}
\end{equation}
where $\Omega_{\rm K}=\left(GM_{\rm NS}/R^{3}\right)^{1/2}$ is the Keplerian velocity,
$n(\omega)=1-\omega$ is the fiducial dimensionless torque which depends on the
fastness parameter $\omega=\Omega_{\rm s}/\left(GM_{\rm NS}/r_{\rm m}^3\right)^{1/2}=\left(r_{\rm m}/r_{\rm c}\right)^{3/2}$.
The latter two torques are described by
\begin{eqnarray}
N_{\rm dip}=\frac{B_{\rm s}^2R^6\Omega_{\rm s}^3}{6c^3},
\label{eq:N_dip}
\end{eqnarray}
and \citep{spr99}
\begin{equation}
N_{\rm mag}=\frac{2}{3}R_{\rm c}^3B_{\rm r}B_{\phi},
\label{eq:N_mag}
\end{equation}
where the radial magnetic field $B_{\rm r}=B_{\rm s}/\epsilon$ and the ratio $\epsilon\simeq0.3$ is given.
The radial magnetic field $B_{\rm r}$ can be linked to the toroidal field $B_{\phi}$ through \citep{spr99}
\begin{equation}
\frac{d B_{\phi}}{d t}=(\Delta\Omega)B_{\rm r}\equiv(\Omega_{\rm s}-\Omega_{\rm c})B_{\rm r}.
\label{eq:dB_phi}
\end{equation}
Besides, the radiation in connection with the magnetic dipole torque is
\begin{eqnarray}
L_{\rm dip}=N_{\rm dip}\Omega_{\rm s}=\frac{B_{\rm s}^2R^6\Omega_{\rm s}^4}{6c^3}.
\label{eq:L_dip}
\end{eqnarray}

To obtain a more realistic expression for the evolution of the accretion rate,
we fit the accretion evolution in low entropy and efficient wind of Figure 3 of \cite{kal22} using a polynomial function and get
\begin{eqnarray}
\dot{M}&\simeq10^{-1.34+0.00167\times {\rm log}_{10}(t)-1.025\times [{\rm log}_{10}(t)]^2}  \nonumber \\
	&\times10^{0.64\times [{\rm log}_{10}(t)]^3-0.47\times [{\rm log}_{10}(t)]^4}\ M_{\odot}{\rm s^{-1}},
\label{eq:M_dot}
\end{eqnarray}
where $t$ is measured in seconds.
Combining with Equations (\ref{eq:r_m})-(\ref{eq:M_dot}), we can perform a numerical calculation
for the angular velocities of the shell and the core, the toroidal field amplification, and the magnetic dipole radiation
via the 4-order Runge-Kutta algorithm.
Note that we have used: (1) initial conditions: initial period of the shell $P_{\rm s,i}=P_{\rm s,0}=1~{\rm s}$, $\Omega_{\rm s,0}=2\pi/P_{\rm s,i}=\Omega_{\rm c,0}$, $B_{\phi,0}=10^0~{\rm G}$, and $B_{\rm s}=10^{14}-10^{15}~{\rm G}$; (2) boundary conditions: $\Omega_{\rm s}\leqslant \Omega_{\rm s,break}=6541~{\rm rad~s^{-1}}$ due to an adopted break-up limit $P_{\rm s,break}=0.96~{\rm ms}$, $\Omega_{\rm c}\leqslant\Omega_{\rm c,max}=c/R_{\rm c}$, and $B_{\phi}\leqslant B_{\rm b}=10^{17}~{\rm G}$ because of the buoyancy effect. 
Notice that the surface magnetic field $B_{\rm s}$ adopted here is a magnetar-like surface field. This is because if the quasi-periodically oscillating precursor of GRB 211211A with a very high peak luminosity $7.4\times10^{49}~{\rm erg~s^{-1}}$ \citep{xiao22} arises from a giant flare of the pre-merger NS, the NS should possess a magnetic field at least a few times $10^{14}$ G \citep{suv22}. Moreover, the surface field strength cannot be $>10^{15}~{\rm G}$ because it would be $r_{\rm m}<r_{\rm c}$ (Equations (\ref{eq:r_m}) and (\ref{eq:r_c})) and the accretion will be into the ``propeller regime" after $\sim13$ s if $B_{\rm s}>10^{15}~{\rm G}$. Thus, to power the duration $\sim$13 s of the MB seen in GRB 211211A by accretion, $B_{\rm s}\lesssim10^{15}~{\rm G}$ is required.

\begin{figure}
\includegraphics[width=0.5\textwidth, angle=0]{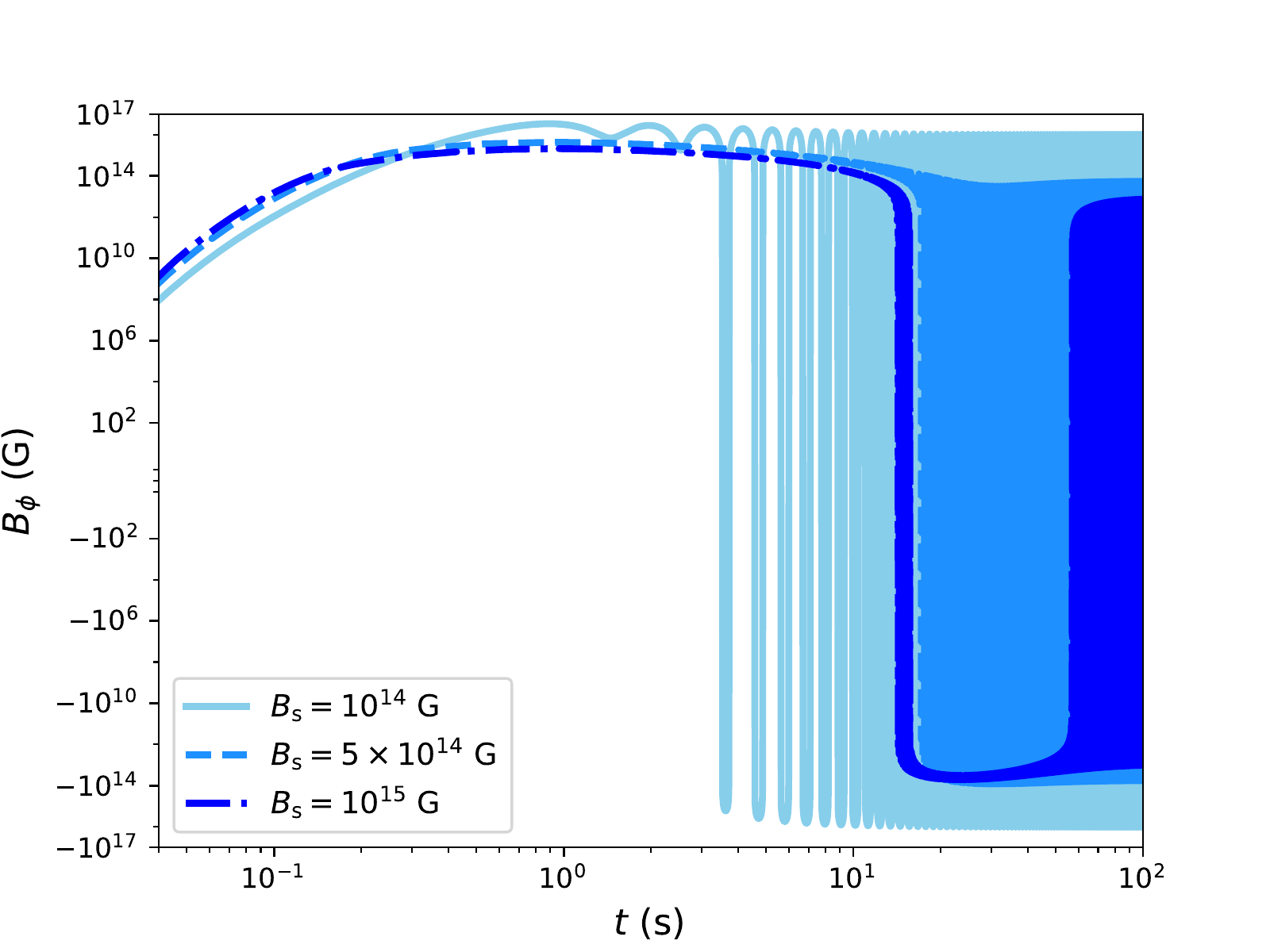}
\includegraphics[width=0.5\textwidth, angle=0]{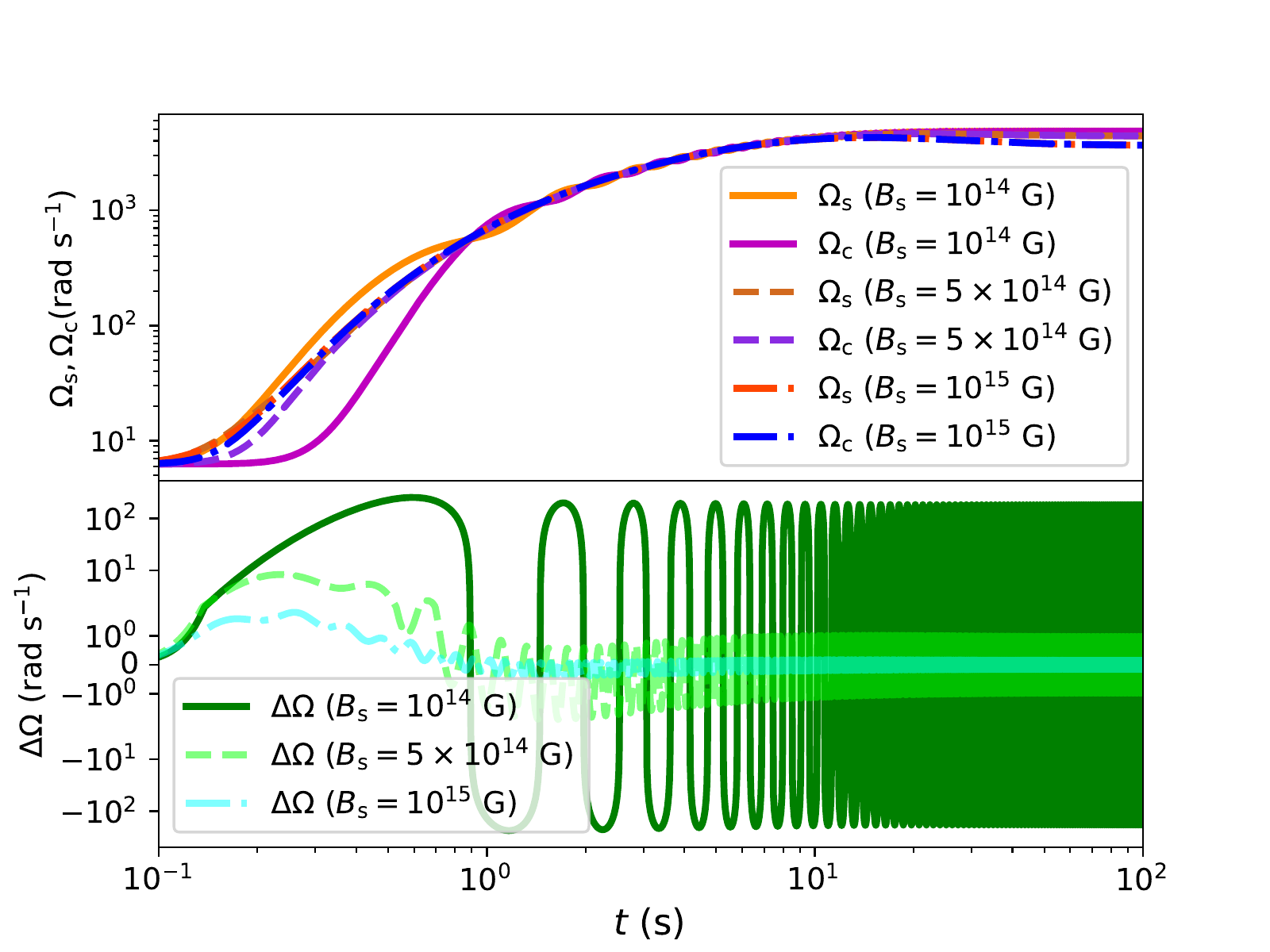}
\includegraphics[width=0.5\textwidth, angle=0]{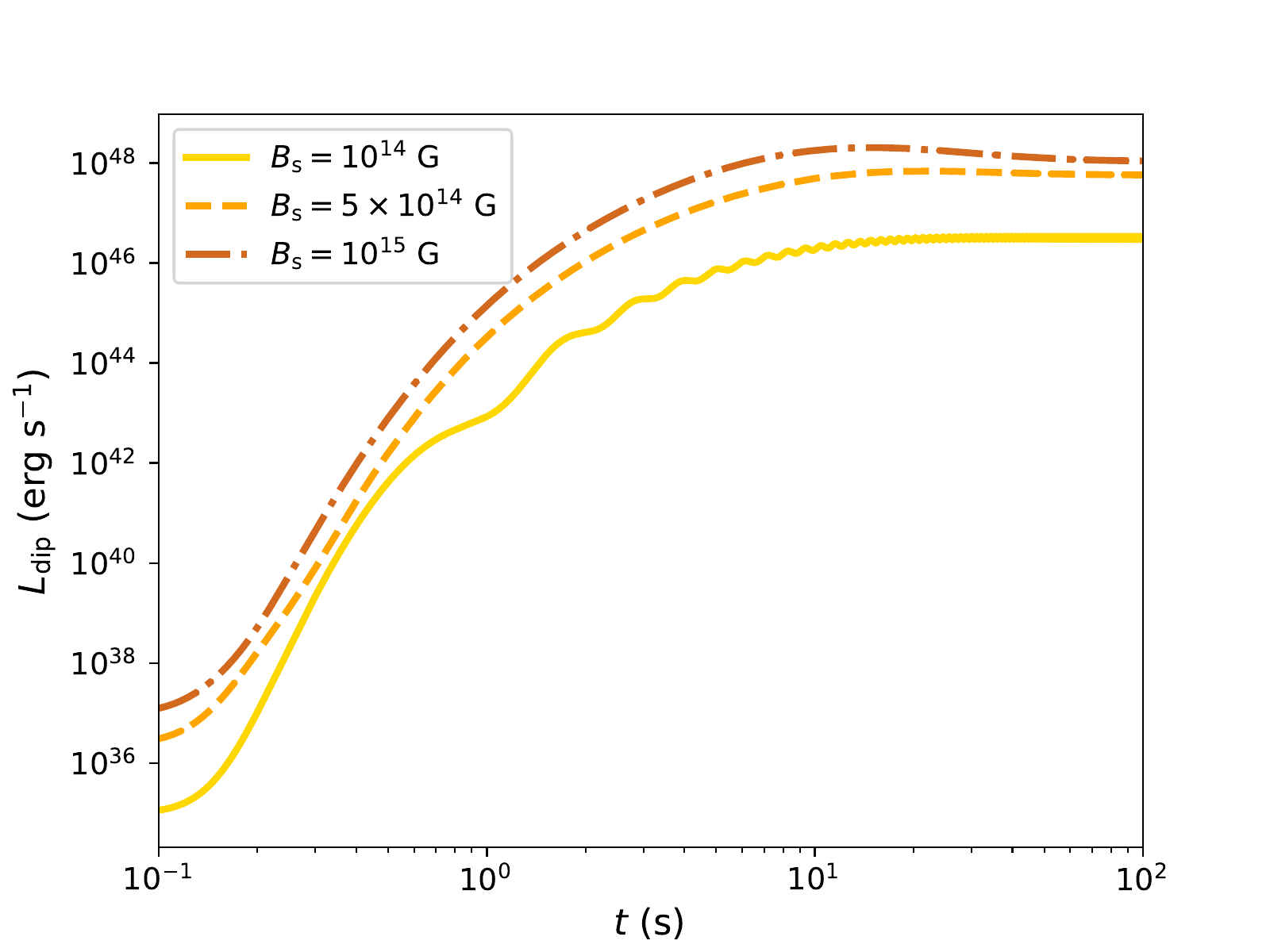}
\caption{Evolutions of the torodial magnetic field $B_{\phi}$ (upper panel), the angular velocities of the shell $\Omega_{\rm s}$ and the core $\Omega_{\rm c}$ as well as $\Delta\Omega=\Omega_{\rm s}-\Omega_{\rm c}$ (middle panel), and the luminosity of the magnetic dipole radiation $L_{\rm dip}$ (bottom panel) of the NS in an NS$-$WD merger during the accretion timescale.}
\label{fig1}
\end{figure}

The final numerical results are exhibited in Figure \ref{fig1} and presented as follows:
\begin{itemize}
\item As shown in the upper panel, the toroidal field $B_{\phi}$ can be quickly enhanced to the peak strength $B_{\phi, \rm peak}\sim5\times10^{15}-10^{17}~{\rm G}$ close to the buoyancy value at $\sim1~{\rm s}$ since the beginning of the accretion. During about one ten to several tens of seconds, the accretion torque in Equation (\ref{eq:I_s}) is dominated and the MB of GRB 211211A should be produced via the transfer from the accretion material to the magnetic bubbles. Then, the accretion will be into the ``propeller regime'', while the toroidal field acts like a resonator because the magnetic torque $N_{\rm mag}$ is gradually dominated, analogous to the result of \cite{dai06}. 
If one once removes the energy dissipation via magnetic bubble, the amplitude of the resonator should decays quickly.
Additionally, it is reasonable to think that the EE of GRB 211211A is produce by the magnetic propellering mechanism \citep{gom14,gib17} after $>13$ s. From the aspect of energetics and duration, its peak luminosity $\sim2\times10^{50}~{\rm erg~s^{-1}}$ and the main pulse of the EE lies in the time range of $\sim15-30$ s. 
One may use Equation (\ref{eq:L_j}) to roughly estimate that it requires the accretion (actually the propeller) rate $>10^{-4}-10^{-3}M_{\odot}{\rm s}^{-1}$. 
This can be meet if the disk is low-entropy and efficient wind in the time range of $\sim15-30$ s, see Figure 3 of \cite{kal22}. The later weak emission of the EE is corresponding to the decline accretion from $10^{-4}M_{\odot}{\rm s}^{-1}$ to $10^{-6}M_{\odot}{\rm s}^{-1}$, lying in the time range of $\sim30-60$ s.
Accordingly, 
the explanation of the MB of GRB 211211A resulting from the accretion and the EE from the propeller is possible in this model. 
Furthermore, the transition from accretion to propeller can naturally explain the trough between the MB and EE in GRB 211211A. 
This may be an important signature to distinguish our model from others.
\item  Not only $B_{\phi}$ can be amplified, but also both the shell and core of the NS can be spun up to the smallest period $P_{\rm s,\min}\sim1.3-1.5$ ms at $>13~{\rm s}$, 
then gradually and slightly spun down if $B_{\rm s}\sim10^{14}-10^{15}~{\rm G}$ (the middle panel).
Meanwhile, the magnetic dipole radiation also increases up to its maximal value $L_{\rm dip,peak}\sim10^{46}-10^{48}$ erg at $>13~{\rm s}$ then slowly decays (the bottom panel).
This dipole radiation could be important as energy injection into both the ejecta producing a kilonova-like emission and the jet generating an X-ray plateau in afterglows, see Section \ref{sec:kilo-after}.
\end{itemize}

\section{Kilonova-like Emission and Afterglows}
\label{sec:kilo-after}
\subsection{Kilonova-like Emission}
\label{subsec:kilonova}
Observationally, a class of RETs are possibly relevant to an NS$-$WD merger
such as SN 2005ek \citep{drout13}, SN 2010X \citep{kasli10}, SN 2018kzr \citep{mcb19,gill20},
and SN 2019bkc \citep{chen20,pren20}.
Theoretically, this possible relation between some RETs and NS$-$WD mergers is also supported by the hydrodynamical thermonuclear simulations for the mass ejection from the accretion discs formed in NS$-$WD mergers \citep{met12,fern13,mar16,fern19,zen19,zen20}.
Interestingly, several RETs like SN 2005ek \citep{drout13} and SN 2018kzr \citep{mcb19} require an extra energy source besides the radioactive power by $^{56} \rm{Ni}$ decay.
The kilonova-like emission accompanying GRB 211211A to some extent is in line with those RETs invoking an NS$-$WD merger and requiring an extra energy source, since the model only containing the pure $^{56} \rm{Ni}$ power cannot well fit the kilonova-like emission \citep{ras22}.
If the final NS is a magnetar after an NS$-$WD merger,
as explored in Section \ref{sec:prompt},
the required extra energy source can be the magnetic dipole radiation of the magnetar [Equation (\ref{eq:L_dip})].

\begin{figure*}
\includegraphics[width=1.\textwidth, angle=0]{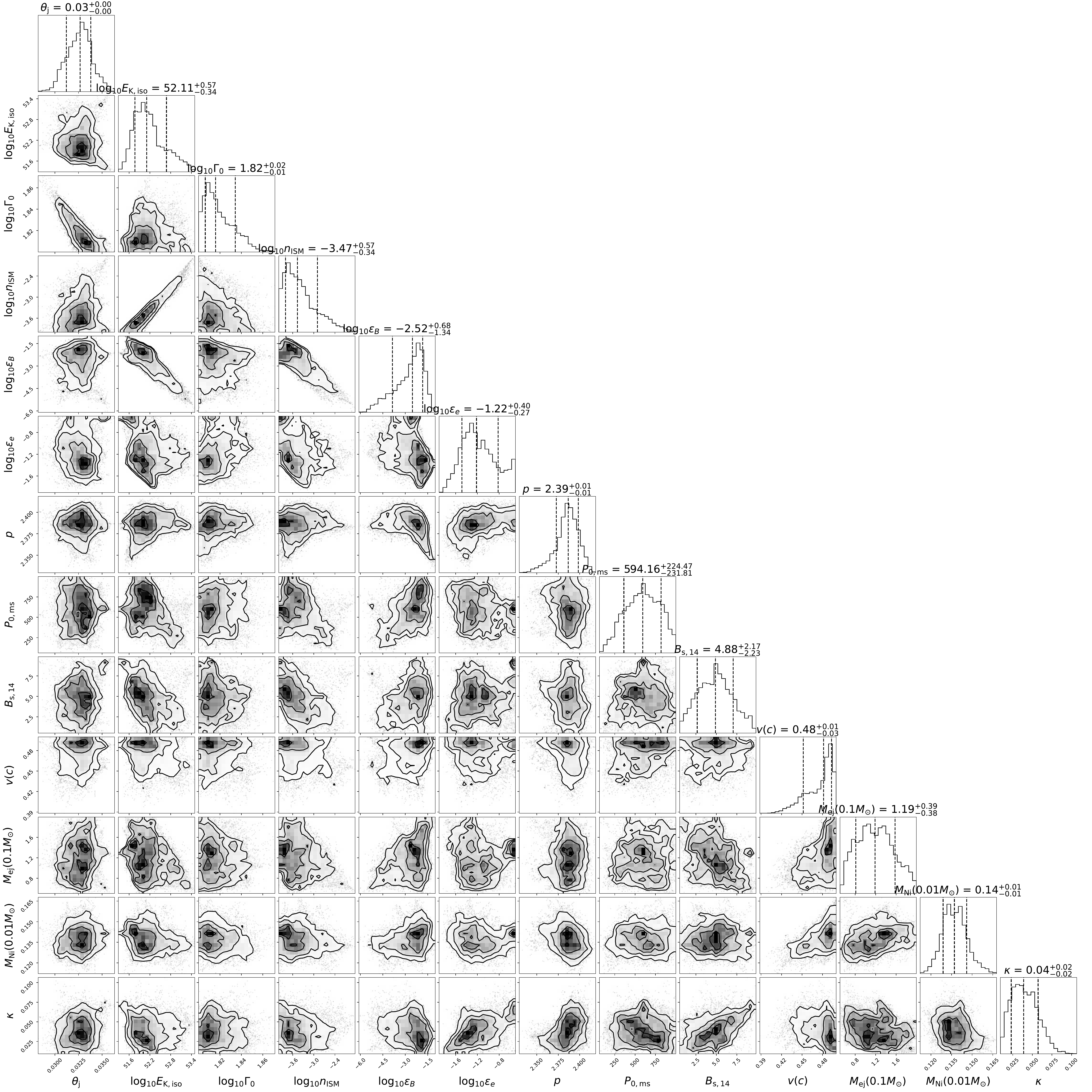}
\caption{The corner plot of parameters for jointly modeling the kilonova-like emission and afterglows by the MCMC algorithm:
the jet half-opening angle $\theta_{\rm j}$, the isotropic kinetic energy $E_{\rm K,iso}$,
the initial Lorentz factor $\Gamma_0$,
the ratio of shock energy to the magnetic field $\epsilon_B$, the ratio of shock energy to the electron $\epsilon_e$, the medium density $n$,
the power-law index of the electron distribution $p$, the initial period $P_0$ and surface magnetic field $B_{\rm s}$ of the magnetar, the expansion velocity $v$ and mass $M_{\rm ej}$ of the ejecta, the mass of $^{56} \rm{Ni}$, and the opacity $\kappa$. The former seven are involved with the standard afterglow model, and the rest with the $^{56}{\rm Ni}$ decay power plus a magnetar spin-down power.}
\label{fig2}
\end{figure*}

Consequently, we model the kilonova-like emission using a semi-analytical formula
for the bolometric luminosity from the radioactive power of
$^{56} \rm{Ni}$ decay plus a power from the magnetic dipole radiation,
as employed in SNe \citep[e.g.,][]{kasen10,woo10,chatz12,inserra13,wang15a,wang15b,lilong20} and RETs \citep{mcb19},
\begin{equation}
	\begin{aligned}
		L(t)=& \frac{2}{\tau_{m}} e^{-\frac{t^{2}}{\tau_{m}^{2}}}  \\
		\times& \int_{0}^{t} e^{\frac{t^{\prime 2}}{\tau_{m}^{2}}} \left(\frac{t^{\prime}}{\tau_{m}}\right) P(t^{\prime}) d t^{\prime}\ \mathrm{erg~s}^{-1},
	\end{aligned}
\label{eq:Lt}
\end{equation}
where $\tau_{m}$ is the diffusion timescale parameter, given by
\begin{equation}
	\tau_{m}=\left(\frac{2 \kappa M_{\rm ej}}{\beta v c}\right)^{1 / 2},
	\label{eq:tau_m}
\end{equation}
in which $\kappa$ is the opacity to optical photons, $M_{\rm ej}$ is the ejecta mass, $v$ is the ejecta expansion velocity,
$\beta$ is a parameter accounting for the
density distribution of the ejecta with a typical value 13.8 \citep{arnett82}.
The total power $P(t)$ is written as
\begin{equation}
P(t)=P_{\rm Ni}(t)+P_{\rm NS}(t),
	\label{eq:Pt}
\end{equation}
where the first term is the radioactive power of $^{56}{\rm Ni}$ and its daughter nucleus $^{56}{\rm Co}$ decay
\begin{equation}
	P_{\rm Ni}(t)=\epsilon_{\rm Ni} M_{\rm Ni} e^{-t / \tau_{\rm Ni}}+\epsilon_{\rm Co} M_{\rm Ni}
	\frac{e^{-t / \tau_{\rm Co}}-e^{-t / \tau_{\rm Ni}}} {1-\tau_{\rm Ni} / \tau_{\rm Co}},
	\label{eq:P_Ni}
\end{equation}
where $M_{\rm Ni}$ is the amount of $^{56}{\rm Ni}$ formed in the explosion, $\tau_{\rm Ni}=8.8~{\rm days}$ and $\tau_{\rm Co}=111.3~{\rm days}$ are the decay times of the isotopes, $\epsilon_{\rm Ni}=3.9 \times 10^{10}~{\rm erg~s^{-1}~g^{-1}}$ \citep{suther84,capp97}
and $\epsilon_{\rm Co}=6.8 \times 10^{9}~{\rm erg~s^{-1}~g^{-1}}$ \citep{maeda03} are their energy generation
rate per unit mass.
The second term comes from the magnetar spin-down,
which is another form of Equation (\ref{eq:L_dip}),
\begin{equation}
	P_{\rm NS}(t)=\frac{E_{\mathrm{NS}}}{\tau_{\mathrm{NS}}} \left(1+\frac{t}{\tau_{\rm NS}}\right)^{-2},
	\label{eq:P_ns}
\end{equation}
where $E_{\mathrm{NS}}=\frac{1}{2} I \Omega_{0}^{2}$ is the rotational energy of the magnetar
linked to its initial spin period
$P_0=\frac{2\pi}{\Omega_{0}}$,
$\tau_{\mathrm{NS}}=\frac{3 I c^{3}}
{B_{\rm s}^{2} R^{6} \Omega_{0}^{2}}$ is the characteristic spin-down timescale.

To model the multi-wavelength light curves of the kilonova-like emission, one should reproduce the observed flux density $F_\nu (t)$ in units of ${\rm erg~s^{-1}~cm^{-2}~Hz^{-1}}$ from the bolometric luminosity $L(t)$ and
the photosphere temperature $T(t)$ evolution during the ejecta expansion, through
\begin{equation}
	F_{\nu}(t)=\frac{2 \pi h \nu^{3}}{c^{2}} \frac{1}{e^{h \nu / k T}-1}\frac{v^2t^2}{D_{\rm L}^2},
	\label{eq:F_v}
\end{equation}
and
\begin{equation}
	T(t)=\left[\frac{L(t)}{4\pi \sigma_{\rm SB} v^2 t^2}\right]^{1/4},
	\label{eq:T}
\end{equation}
where $D_{\rm L}$ is the luminosity distance of the source and $\sigma_{\rm SB}$ is the Stefan-Boltzmann constant.
Finally, one can calculate the observed monochromatic
AB magnitude through $M_\nu(t)=-2.5\times{\rm log_{10}}[F_{\nu}(t)/3631{\rm Jy}]$.

\begin{deluxetable}{ccccc}
\label{tab:parameters}
\tablecaption{The Best Parameter Distributions for Jointly Modeling the Kilonova-like Emission and Afterglows of GRB 211211A}
\tablehead{
\colhead{Parameters} &
\colhead{Values}
}
\startdata
\object{Afterglow}  &      \\
\hline
\object{$\theta_{\rm j}$ (rad)}  &  $0.033^{+0.002}_{-0.002}$     \\
\object{$\log_{10}(E_{\rm K,iso}/{\rm erg})$} &  $52.110^{+0.570}_{-0.341}$     \\
\object{$\log_{10}(\Gamma_0)$} &  $1.816^{+0.019}_{-0.010}$     \\
\object{$\log_{10}(n/{\rm cm}^{-3}$)}      &  $-3.467^{+0.571}_{-0.338}$     \\
\object{$\log_{10}(\epsilon_B)$}     &    $-2.519^{+0.682}_{-1.337}$   \\
\object{$\log_{10}(\epsilon_e)$}     &    $-1.217^{+0.399}_{-0.267}$   \\
\object{$p$}     &    $2.387^{+0.012}_{-0.014}$   \\
\hline
\object{$^{56}{\rm Ni}$ + post-merger magnetar power}  &     \\
\hline
\object{$P_0$ (ms)}     &    $594.161^{+224.473}_{-231.815}$   \\
\object{$B_{\rm s}$ (G)}     &    $(4.882^{+2.172}_{-2.228})\times10^{14}$   \\
\object{$v$ ($c$)}     &    $0.482^{+0.012}_{-0.030}$   \\
\object{$M_{\rm ej}$ ($M_{\odot}$)}     &    $0.119^{+0.040}_{-0.038}$   \\
\object{$M_{\rm Ni}$ ($M_{\odot}$)}     &    $0.0014^{+0.0001}_{-0.0001}$   \\
\object{$\kappa$ (cm$^2$g$^{-1}$)}     &    $0.039^{+0.019}_{-0.016}$   \\
\enddata
\end{deluxetable}

\begin{figure}
\includegraphics[width=0.5\textwidth, angle=0]{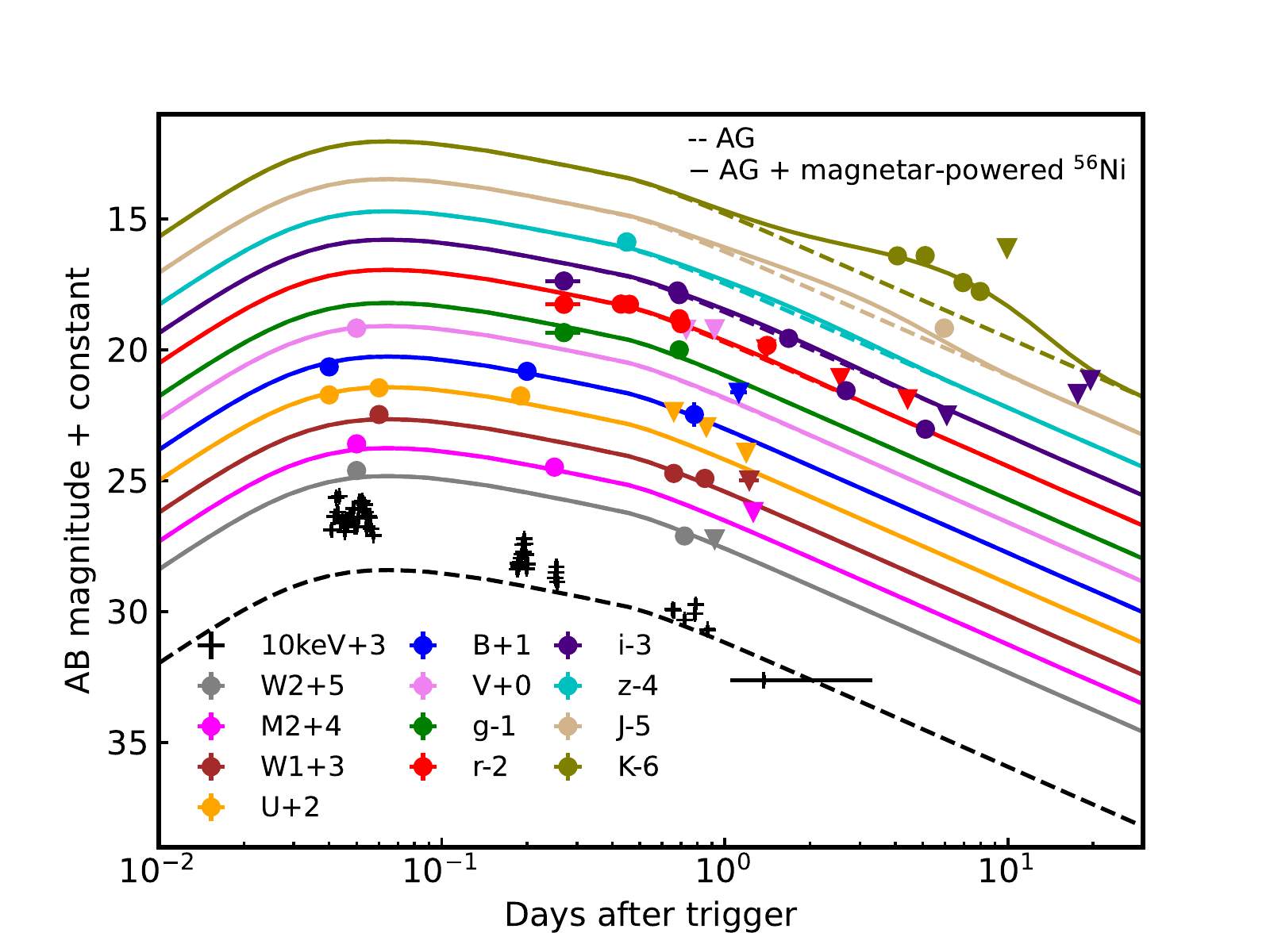}
\caption{The modeling results for the multi-wavelength light curves of the kilonova-like emission and GRB afterglows. The dash lines are the standard afterglow (AG) model, while the solid lines are the standard afterglow model combining with the model of $^{56}{\rm Ni}$ decay power plus a magnetar spin-down power. The data points denoted by an inverted triangle are upper limits.}
\label{fig3}
\end{figure}

\begin{figure}
\includegraphics[width=0.5\textwidth, angle=0]{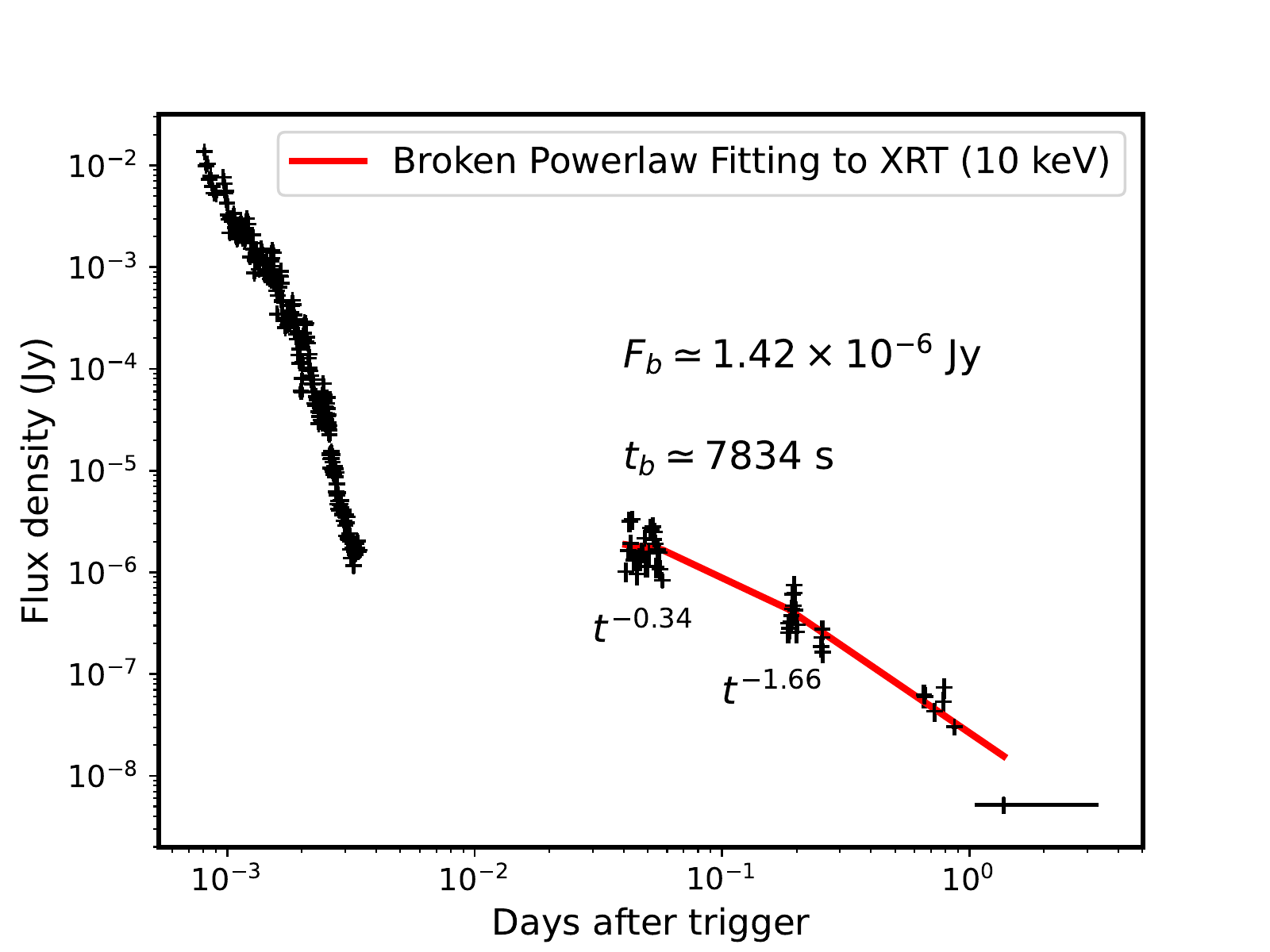}
\caption{The empirical fitting of X-ray afterglow.}
\label{fig4}
\end{figure}

\subsection{Jointly Modeling the Kilonova-like Emission and Afterglows}
\label{subsec:joint}
It is widely accepted that the multi-wavelength afterglows of a GRB are generated by the synchrotron radiation within the external shocks due to the collisions between the GRB jet and its surrounding medium.
We use the standard external shock model \citep{sari98,huang99} to reproduce the observed multi-wavelength afterglows of GRB 211211A.

Owing to the observed optical, near-infrared data collected from \cite{ras22} and the Swift UVOT/XRT light curves \citep{rom05,evans09} should contain both the kilonova-like emission and GRB afterglows,
we process a joint fit to the observed data combining the magnetar-fed $^{56}{\rm Ni}$ power model (Section \ref{subsec:kilonova}) with the standard afterglow model.
The free parameters are the jet half-opening angle $\theta_{\rm j}$, the isotropic kinetic energy $E_{\rm K,iso}$,
the initial Lorentz factor $\Gamma_0$,
the ratio of shock energy to the magnetic field $\epsilon_B$, the ratio of shock energy to the electron $\epsilon_e$, the medium density $n$,
the power-law index of the electron distribution $p$, the initial period $P_0$ and surface magnetic field $B_{\rm s}$ of the post-merger magnetar, the expansion velocity $v$ and mass $M_{\rm ej}$ of the ejecta, the mass of $^{56} \rm{Ni}$, and the opacity $\kappa$.
To search for the best parameter values to fit the data, we run the Markov chain Monte Carlo (MCMC) algorithm within reasonable prior ranges of parameter values. 
Some meaningful results are obtained as follows:
\begin{itemize}
\item The best parameter set are shown in Figure \ref{fig2} and Table \ref{tab:parameters}.
The values for the mass $M_{\rm ej}$, velocity $v$, and opacity $\kappa$ of the ejecta, 
and the $^{56}{\rm Ni}$ mass all lie in the reasonable ranges from the recent hydrodynamic simulations
for an NS$-$WD binary with mass ratio $M_{\rm NS}:M_{\rm WD}=1.40:1.00$ \citep{kal22}. 
The values of the standard afterglow model parameters are also located in the typical ranges of SGRBs. With this parameter set, the multi-wavelength
light curves of the kilonova-like emission and afterglow are well reproduced, see Figure \ref{fig3}. 
The goodness of the fitting is not worse than the best-fits in \cite{ras22}, \cite{yang22}, and \cite{tro22}.
\item There seems to be a plateau in X-ray afterglow exhibited in Figure \ref{fig3}. Also see Figure \ref{fig4}, the break time $t_{\rm b}\simeq7834$ s, 
the break flux density $F_{\rm b}\simeq1.42\times10^{-6}$ Jy, and the post-break slope $\simeq-1.66$ can be obtained from the empirical broken power-law fitting to the X-ray data.
Thus, the characteristic break timescale in the source frame $\tau=t_{\rm b}/(1+z)$ ($z$ is the redshift), 
and the break luminosity $L_{\rm b}=4\pi D_{\rm L}^2F_{\rm b}$ are thus acquired.
This plateau and its subsequent decay with a slope close to $-2$ is usually thought to be the signature of magnetar wind learned from previous classical works \citep[e.g.,][]{dai98a,dai98b,zhang01}. 
If this is the case, one can evaluate the surface magnetic field strength $B_{\rm s,X}\lesssim10^{15}$ G and the initial period  $P_{\rm 0,X}\lesssim2$ ms of the magnetar 
by using the following two equations \citep{zhang01}
\begin{eqnarray}
	B_{\rm s,X}&=&\sqrt{\frac{3}{2}}c^{3/2} I R^{-3} L_{\rm X,iso}^{-1/2} \tau^{-1} \nonumber\\
	&\simeq& 2.24\times10^{17}\ M_{\rm NS,1.4} R_{6}^{-1} L_{\rm X,iso,45}^{-1/2} \tau_3^{-1}\ {\rm G},
	\label{eq:B_sX}
\end{eqnarray}
and
\begin{eqnarray}
	P_{\rm 0,X}&=&\sqrt{2}\pi I^{1/2} L_{\rm X,iso}^{-1/2} \tau^{-1/2} \nonumber\\
	&\simeq& 0.15\ M_{\rm NS,1.4}^{1/2} R_{6} L_{\rm X,iso,45}^{-1/2} \tau_3^{-1/2}\ {\rm s},
	\label{eq:P_0X}
\end{eqnarray}
where $\tau_3=\tau/10^3{\rm s}$, $L_{\rm X,iso,45}=L_{\rm X,iso}/10^{45}{\rm erg~s^{-1}}$, and $L_{\rm X,iso}=L_{\rm b}/\eta_{\rm X}$ with a relatively low radiative efficiency $\eta_{\rm X}\lesssim10^{-3}$ \citep{xiao19}.
These values of surface magnetic field and initial period of the post-merger magnetar are exactly satisfied after the accretion of the pre-merger NS from the WD material, see the middle and bottom panels of Figure \ref{fig1} and the discussion in the last paragraph of Section \ref{sec:prompt}.
\item The jet half-opening angle from our modeling is $\theta_{\rm j}\sim0.033~{\rm rad}\sim2^{\circ}$, which is consistent with that from other papers \citep{ras22,yang22,tro22,mei22}
and is slightly smaller than the jet core half-opening angle $\sim3.4^{\circ}$ in GRB 170817A \citep{ghir19}. Moreover, the jet break time mainly obtained from our joint modeling for optical afterglows and kilonova-like emission is at $\sim0.4-0.6$ days, see Figure \ref{fig3}.
\item The initial period of the post-merger magnetar inferred from the kilonova-like emission and afterglow modeling is $P_0\sim594$ ms, 
while that from the X-ray plateau fitting is $P_{\rm 0,X}\lesssim2$ ms, 
this large discrepancy could be attributed to the structured distribution of the magnetar wind radiation. That is, one part of the magnetar wind radiation is beamed into the jet to dissipate as the X-ray plateau, while the other part is isotropic and poured into the ejecta that gives rise to the kilonova-like emission. The structured distribution of the magnetic spin-down power is just like 
\begin{equation}
	L_{\rm sd}(\theta)=\begin{cases}L_{\rm X}=L_{\rm X,iso}f_{\rm b}\gtrsim3\times10^{44}~{\rm erg~s^{-1}}, &\theta\lesssim \theta_{\rm j}, \\
		E_{\rm NS}/\tau_{\rm NS}\sim2\times10^{37}~{\rm erg~s^{-1}}, &\theta> \theta_{\rm j},
	\end{cases}
	\label{eq:L_sd}
\end{equation}
where $f_{\rm b}\simeq0.5\theta_{\rm j}^2$ is the beaming factor, $E_{\rm NS}$ and $\tau_{\rm NS}$ are referred to Equation (\ref{eq:P_ns}).
\item From our modeling results in Table \ref{tab:parameters}, the initial Lorentz factor $\Gamma_0$ seems somewhat low, but is comparable to that obtained by \cite{ras22} and located in the wide distribution of GRBs derived by \cite{ghir18}. Meanwhile, the medium density $n$ is also very low, but still falls into the wide range of SGRBs \citep{fong15}.

\end{itemize}

\section{Event Rate}
\label{sec:rate}
There are about one quarter EE-SGRBs in the regular SGRB population \citep{norris10}. 
If the volumetric event rate of SGRBs $\sim270~{\rm Gpc^{-3}~yr^{-1}}$ is adopted \citep{nicholl17b}, 
the rate of the EE-SGRBs is $\sim70~{\rm Gpc^{-3}~yr^{-1}}$, which is much lower than the rate of NS$-$WD mergers
$(0.5-1)\times10^4~{\rm Gpc^{-3}~yr^{-1}}$ \citep{thom09}. If all EE-SGRBs are thought to originate from NS$-$WD mergers, the rate comparison indicates that only $\sim1$\% can form EE-SGRBs. This is because only a small fraction of NS$-$WD mergers can fulfill: (1) the WD with a required mass $\gtrsim1M_{\odot}$ in NS$-$WD mergers, and (2) an EE-SGRB like GRB 211211A with the requirement that the NS in NS$-$WD mergers is a magnetar. From the demographics study of NS$-$WD mergers using binary population synthesis models for the isolated primordial binary channel by \cite{too18}, there are a few times 0.1\% mergers in which the SN producing the NS precedes the NS$-$WD merger by less than 100 years. These young NSs in this fraction of NS$-$WD mergers possibly are magnetars. Therefore, it seems the rate of NS$-$WD mergers with a pre-merger magnear is somewhat lower than that fraction producing EE-SGRBs. However, it is still uncertain whether the fast evolution channels such as formation in clusters or in the field through dynamical captures are needed. This is because: (1) whether all EE-SGRBs invoke an NS$-$WD merger with a pre-merger magnetar is unclear, and (2) whether the NS$-$WD mergers having a pre-merger magnetar likewise having a WD with mass $\gtrsim1M_{\odot}$ is also unknown.

Alternatively, if all SGRBs with or without EE stem from NS$-$NS and NS$-$BH mergers, one can see that the event rate of EE-SGRBs is also roughly comparable to the quarter of the total event rate of NS$-$NS and NS$-$BH mergers $\sim20-1100~{\rm Gpc^{-3}~yr^{-1}}$ \citep{ligo21,abb21}. But for EE-SGRBs like GRB 211211A, they require the pre-merger NS is a magnetar, these cases might invoke an NS$-$NS  or NS$-$BH binary formation via a fast dynamical formation channel of binary evolution such as young star cluster or active galactic nuclei channel.

\section{Summary and Discussion}
\label{sec:summary}
GRB 211211A and its associated kilonova-like emission reported recently may be the second confirmed association between GRB and kilonova from
an NS$-$NS merger.
In this paper, however, we proposed another model to explain the prompt emission, afterglows, and its kilonova-like emission of GRB 211211A within the framework of an NS$-$WD
merger. In this model, on one hand, the toroidal magnetic field of the pre-merger NS can be amplified by the runaway material accretion from the disrupted WD if the WD disc is in low initial entropy and efficient wind. When the field reaches up to the buoyancy limit and would give rise to magnetic bubble eruptions. 
These magnetic bubble eruptions lead to the MB of GRB 211211A. 
After the MB, the accretion is into the propeller regime and the WD materials power the EE of GRB 211211A via the magnetic propelling.
On the other hand, during the accretion, part of the WD debris burns and would throw out an ejecta, 
this ejecta with the $^{56}{\rm Ni}$ decay power fed by a part of post-merger magnetar rotational power results in the kilonova-like emission.
The other part of the magnetar rotational power pours into the jet and dissipates as the X-ray plateau in afterglows.
Notice that the post-merger magnetar rotational power injected into the ejecta and the jet actually comes from the spin-up of the pre-merger NS during the accretion.

We have showed that the prompt emission, afterglows, and its kilonova-like emission of GRB 211211A can be well explained in this model. Some results and notes should be emphasized as follows:
\begin{itemize}
\item The MB is created by the magnetic bubble eruptions, which stem from the amplification of the toroidal magnetic field $B_{\phi}$ in the pre-merger NS owing to the material accretion from the WD. While the EE is produced by the subsequent accretion but via the magnetic propelling mechanism since the accretion has been into the propeller regime. The transition from accretion to propeller can naturally account for the trough between the MB and EE in GRB 211211A. 
This may be an important signature to distinguish our model from others. Comparing the accretion characteristics in which the disc is in low initial entropy and efficient wind, with those of the prompt emission of GRB 211211A, one can get that a massive WD $M_{\rm WD}\gtrsim1M_{\odot}$ in the NS$-$WD merger is required from the hydrodynamical thermonuclear simulation of \cite{kal22}.
\item The kilonova-like emission and optical afterglows can be well modeled by the radioactive decay power of $^{56}{\rm Ni}$ adding a part of magnetar spin-down power. The other part of the magnetar spin-down power is poured into the jet to generate the X-ray plateau in afterglows.
\end{itemize}

Overall, GRB 211211A and its kilonova-like emission seem to be well accounted for either in an NS$-$WD, NS$-$NS, or NS$-$BH merger under special conditions, suggested in this paper and previous works. This signifies that it is difficult to firmly identify the origin of a GRB 211211A-like case through its prompt emission, afterglows, and multi-wavelength kilonova-like emission.
Hence, a further way like presenting a detailed spectroscopic analysis for spectral line features should be needed, which can help with diagnosing the element abundance in ejecta material, as done for AT 2017gfo \citep{wat19} and SN 2018kzr \citep{gill20}.

Additionally, the pre-merger and post-merger GW emissions may be another point to identify the origin of a GRB 211211A-like case. 
As if its progenitor is an NS$-$WD merger, there will be no inspiral GW emission with frequency into the LIGO-Virgo-KAGRA (LVK) detection range prior to the merger. 
This is different from an NS$-$NS or NS$-$BH merger. \cite{sarin23} showed that a 1.4 + 4.5 $M_{\odot}$ NS$-$BH merger (dimensionless spin $a = 0.85$)
and a 1.4 + 1.4 $M_{\odot}$ NS$-$NS merger at 350 Mpc with GRB 211211A-like electromagnetic
observations can be detectable in the LVK O4 running during inspiral phase. 
Nevertheless, the post-merger GW seems cannot directly distinguish the progenitor. 
Because, on one hand, it either is an NS$-$WD or NS$-$NS, a BH remnant is possible, similar to an NS$-$BH merger.
On the other hand, whether it is an NS$-$WD or NS$-$NS merger, a possible millisecond magnetar leaves behind and may lead to strong high-frequency GW radiation. 
Taking GRB 211211A as an instance, if the GW radiation of the post-merger magnetar dominates its spin-down, the characteristic GW strain can be given by  \citep{corsi09,lasky16,lv17}
\begin{eqnarray}
h_{\mathrm{c}} &=&\frac{f}{D_{\mathrm{L}}} \sqrt{\frac{5 G I}{2 c^3 f}} \nonumber \\
&\approx& 8.22 \times 10^{-24}\left(\frac{I}{10^{45} \mathrm{~g} \mathrm{~cm}^2} \frac{f}{1 \mathrm{kHz}}\right)^{1 / 2}\left(\frac{D_{\mathrm{L}}}{100 \mathrm{Mpc}}\right)^{-1},
\end{eqnarray}
where $f=2/P_{\rm s,\min}$. Using $P_{\rm s,\min}\sim1.5~{\rm ms}$ and $D_{\rm L}=347.8~{\rm Mpc}$, $h_{\rm c}\sim2.73\times10^{-24}$ is above the threshold of Einstein Telescope but below LVK. 
In short, the pre-merger GW detection by LVK for a nearby EE-SGRB like GRB 211211A maybe enough to identify the origin of the EE-SGRB.

\acknowledgments
We are very grateful to thank the referee for her/his careful and thoughtful suggestions that have helped improve this manuscript substantially.
S.Q.Z. and L.L. acknowledge support from the National Natural Science Foundation
of China (grant No. 12247144) and China Postdoctoral Science Foundation (grant Nos. 2021TQ0325 and 2022M723060).
Z.G.D. is supported by the National Key
Research and Development Program of China (grant No.
2017YFA0402600), National SKA Program of China (grant No. 2020SKA0120300),
and National Natural Science Foundation
of China (grant No. 11833003).
We also acknowledge our use of public data from the {\em Swift} and {\em Fermi} data archive.



\end{document}